\documentclass[aps,prd,twocolumn,showpacs,amsmath,amssymb,preprintnumbers]{revtex4-2}
\usepackage{amsmath}
\usepackage{graphicx}
\usepackage{subfigure}
\usepackage{epstopdf}
\usepackage{color}
\usepackage{multirow}
\usepackage{setspace}
\usepackage{overpic}
\usepackage{amssymb}
\usepackage{makecell}
\usepackage{ulem}
\usepackage[bookmarksnumbered, pdfstartview=FitH,colorlinks,urlcolor=blue, citecolor=blue,linkcolor=blue] {hyperref}
\usepackage{lineno}
\usepackage{bm}
\usepackage{float}
\usepackage{rotating}
\usepackage[utf8]{inputenc}
\usepackage[separate-uncertainty=true,table-align-uncertainty=true,]{siunitx}

\newcommand{\BESIIIorcid}[1]{\href{https://orcid.org/#1}{\hspace*{0.1em}\raisebox{-0.45ex}{\includegraphics[width=1em]{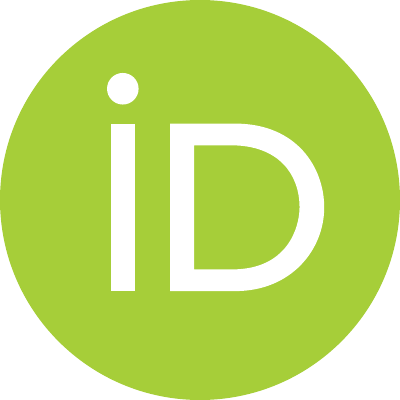}}}}

\let\oldequation\equation
\let\oldendequation\endequation
\renewenvironment{equation}
 {\linenomathNonumbers\oldequation}
 {\oldendequation\endlinenomath}
\begin{document}

\title{\boldmath Study of the Magnetic Dipole Transition of $J/\psi\to\gamma\eta_c$ via $\eta_c\to p\bar{p}$}
\author{
M.~Ablikim$^{1}$\BESIIIorcid{0000-0002-3935-619X},
M.~N.~Achasov$^{4,b}$\BESIIIorcid{0000-0002-9400-8622},
P.~Adlarson$^{77}$\BESIIIorcid{0000-0001-6280-3851},
X.~C.~Ai$^{82}$\BESIIIorcid{0000-0003-3856-2415},
R.~Aliberti$^{36}$\BESIIIorcid{0000-0003-3500-4012},
A.~Amoroso$^{76A,76C}$\BESIIIorcid{0000-0002-3095-8610},
Q.~An$^{73,59,\dagger}$,
Y.~Bai$^{58}$\BESIIIorcid{0000-0001-6593-5665},
O.~Bakina$^{37}$\BESIIIorcid{0009-0005-0719-7461},
Y.~Ban$^{47,g}$\BESIIIorcid{0000-0002-1912-0374}
% ...
\\
\vspace{0.2cm}
(BESIII Collaboration)\\
\vspace{0.2cm} {\it
$^{1}$ Institute of High Energy Physics, Beijing 100049, People's Republic of China\\
% ...
$^{4}$ Budker Institute of Nuclear Physics SB RAS (BINP), Novosibirsk 630090, Russia\\
% ...
$^{36}$ Johannes Gutenberg University of Mainz, Johann-Joachim-Becher-Weg 45, D-55099 Mainz, Germany\\
$^{37}$ Joint Institute for Nuclear Research, 141980 Dubna, Moscow region, Russia\\
% ...
$^{47}$ Peking University, Beijing 100871, People's Republic of China\\
% ...
$^{58}$ Southeast University, Nanjing 211100, People's Republic of China\\
$^{59}$ State Key Laboratory of Particle Detection and Electronics, Beijing 100049, Hefei 230026, People's Republic of China\\
% ...
$^{73}$ University of Science and Technology of China, Hefei 230026, People's Republic of China\\
% ...
$^{76}$ University of Turin and INFN, (A)University of Turin, I-10125, Turin, Italy; (B)University of Eastern Piedmont, I-15121, Alessandria, Italy; (C)INFN, I-10125, Turin, Italy\\
$^{77}$ Uppsala University, Box 516, SE-75120 Uppsala, Sweden\\
% ...
$^{82}$ Zhengzhou University, Zhengzhou 450001, People's Republic of China\\
\vspace{0.2cm}
$^{\dagger}$ Deceased\\
% ...
$^{b}$ Also at the Novosibirsk State University, Novosibirsk, 630090, Russia\\
% ...
$^{g}$ Also at State Key Laboratory of Nuclear Physics and Technology, Peking University, Beijing 100871, People's Republic of China\\
% ...
}
}

\preprint{\vbox{ 
    \hbox{Intended for $Phys.Rev.Lett.$}
    \hbox{Draft version: 2.0}
}}

\noaffiliation

\begin{abstract}
Using $(10.087\pm0.044)\times10^9$ $J/\psi$ events collected with the BESIII detector at the $e^+e^-$ BEPCII collider, we present the first amplitude analysis of $J/\psi\to\gamma p\bar{p}$ with the $p\bar p$ invariant mass in the $\eta_c$ mass region $[2.70,3.05]$~GeV/$c^2$. The product branching fraction $\mathcal{B}(J/\psi\to\gamma\eta_c)\times\mathcal{B}(\eta_c\to p\bar{p})$ is determined to be $(2.11\pm0.02_{\rm stat}\pm0.07_{\rm syst})\times10^{-5}$ with precision improved by one order of magnitude. Combining with the product branching fractions $\mathcal{B}(\eta_c\to p\bar{p})\times\mathcal{B}(\eta_c\to \gamma\gamma)$ and $\mathcal{B}(J/\psi\to\gamma\eta_c)\times\mathcal{B}(\eta_c\to \gamma\gamma)$, the branching fractions of $\mathcal{B}(J/\psi\to\gamma\eta_c)$ and $\mathcal{B}(\eta_c\to\gamma\gamma)$ are calculated to be $(2.29\pm0.01_{\rm stat}\pm0.04_{\rm syst}\pm0.18_{\rm opbf})\%$ and $(2.28\pm0.01_{\rm stat}\pm0.04_{\rm syst}\pm0.18_{\rm opbf})\times10^{-4}$, respectively, which are consistent with the latest lattice quantum chromodynamics calculations. Here, opbf is the uncertainty from the other product branching fractions used in the calculation.


\end{abstract}

\maketitle

The transition between heavy quarkonium systems presents an ideal laboratory to investigate the theory of the strong interaction, quantum chromodynamics~(QCD), in both the perturbative and non-perturbative regions. The magnetic dipole~(M1) transition between the two lowest-lying charmonium states, $J/\psi\to\gamma\eta_c$, is of great interest. The predicted transition width in the non-relativistic limit~\cite{Brambilla:2005zw} is found to be significantly larger than the experimental results~\cite{ParticleDataGroup:2024cfk} by a factor of between 2 and 3. Several theoretical studies have attempted to resolve this long-standing puzzle, including dispersion sum rules~\cite{Khodjamirian:1979fa}, QCD sum rules~\cite{Beilin:1984pf}, relativistic quark models~\cite{Ebert:2002pp}, non-relativistic potential models~\cite{Deng:2016stx,Barnes:2005pb}, effective field theories~\cite{Brambilla:2005zw,Pineda:2013lta,Segovia:2021bjb}, light-cone sum rules~\cite{Guo:2019xqa}, and lattice QCD~(LQCD)~\cite{Dudek:2006ej,Dudek:2009kk,Chen:2011kpa,Becirevic:2012dc,Donald:2012ga,Gui:2019dtm,Colquhoun:2023zbc,Meng:2024axn}. However, a significant discrepancy remains between experimental measurements and theoretical predictions. In particular, LQCD calculations are systematically larger than the Particle Data Group~(PDG) average~\cite{ParticleDataGroup:2024cfk} by approximately a factor of two.

Experimental measurements of $\mathcal{B}(J/\psi\to\gamma\eta_c)$ were early reported by CLEO-c~\cite{CLEO:2008pln}, KEDR~\cite{Anashin:2010nr,Anashin:2014wva}, and Crystal Ball~\cite{Gaiser:1985ix} via inclusive hadronic decays of $\eta_c$. Although these works successfully revealed the asymmetric lineshape of $\eta_c$ and the necessity of damping factors, the potential interference between the $\eta_c$ and non-resonant~(NR) amplitudes was either ignored or insufficient considered. This neglect could potentially result in a bias of up to dozens of percent. Later, BESIII presented some measurements of the product branching fraction~(BF) $\mathcal{B}(J/\psi\to\gamma\eta_c)\times\mathcal{B}(\eta_c\to f)$ in various exclusive final states $f$~\cite{BESIII:2012xdg,BESIII:2012gcb,BESIII:2024rex}, but still with the interference being ignored. However, as demonstrated in Ref.~\cite{BESIII:2024rdn}, even if taking into account this interference in one-dimensional fit to the $\eta_c$ mass spectrum, a considerable uncertainty up to dozens of percent is inevitable due to the unknown NR contributions other than $J^{PC}=0^{-+}$. Besides, two solutions with different interference patterns are found to be indistinguishable in Ref.~\cite{BESIII:2024rdn}, which lead to further ambiguities. Therefore, an amplitude analysis, such as that performed in Refs.~\cite{BESIII:2019hek,BESIII:2016tqi}, is highly desired to incorporate all available information into the fit and provide a more reliable description of the interference, hence benefits precise determinations of the BF and resonance parameters. 

The BESIII collaboration recently reported a BF measurement of $\eta_c\to\gamma\gamma$ in $J/\psi\to\gamma\eta_c$~\cite{BESIII:2024rex}. The product $\mathcal{B}(J/\psi\to\gamma\eta_c)\times\mathcal{B}(\eta_c\to \gamma\gamma)$ is in good agreement with the latest LQCD calculations~\cite{Colquhoun:2023zbc,Meng:2021ecs}, while the BF of $\eta_c\to \gamma\gamma$, determined using the BF of $J/\psi\to\gamma\eta_c$ from the PDG~\cite{ParticleDataGroup:2024cfk}, significantly differs from these LQCD calculations and the PDG global fit value~\cite{ParticleDataGroup:2024cfk} by more than 3$\sigma$. Therefore, independent and precise measurements of the BF of $J/\psi\to\gamma\eta_c$ are crucial for clarifying these discrepancies and testing the theoretical models.

Additionally, the hyperfine mass splitting between the $J/\psi~(1^3 S_1)$ and $\eta_c~(1^1 S_0)$ $c\bar{c}$ systems is of importance for our understanding of interquark potential for charmonium system. Various Lattice-QCD calculations~\cite{Follana:2006rc,Burch:2009az,Kawanai:2011jt,Hatton:2020qhk} have been reported, and most of which are in good agreement with each other within small theoretical uncertainties. Hence, precise determination of $\eta_c$ resonance parameters in experiments is highly desired for the validation of relevant calculations.

In this Letter, by analyzing $(10.087\pm0.044)\times10^9$ $J/\psi$ events~\cite{BESIII:2021cxx} collected with the BESIII detector at the symmetric $e^+e^-$ collider BEPCII, we present the first amplitude analysis of $J/\psi\to\gamma p\bar{p}$ with the $p\bar p$ invariant mass $M_{p\bar{p}}$ in the $\eta_c$ mass region $[2.70,3.05]$~GeV/$c^2$, based on which precise measurements on $\eta_c$ contributions and resonance parameters are performed.

Details about the design and performance of the BESIII detector are provided in Refs.~\cite{BESIII:2009fln,Yu:IPAC2016-TUYA01,BESIII:2020nme,etof}. The inclusive Monte Carlo~(MC) sample, which includes both the production of the $J/\psi$
resonance and the continuum processes incorporated in the {\sc kkmc}~\cite{ref:kkmc} generator, are employed to study potential background contributions. All particle decays are modeled with the {\sc evtgen} tool~\cite{ref:evtgen} using branching fractions either taken from the PDG~\cite{ParticleDataGroup:2024cfk}, when available, or otherwise estimated with the {\sc lundcharm} model~\cite{ref:lundcharm}. Final state radiation~(FSR) from charged final state particles is incorporated using {\sc photos}~\cite{photos2}. The simulations of exclusive MC samples are described below.




Candidates for $J/\psi\to\gamma p\bar{p}$ must have two charged tracks with zero net charge. Charged tracks detected in the main drift chamber~(MDC) are required to be within a polar angle ($\theta$) range of $|\rm{cos\theta}|<0.93$, where $\theta$ is defined with respect to the $z$-axis, which is the symmetry axis of the MDC. Their distance of closest approach to the interaction point must be less than 10\,cm along the $z$-axis, and less than 1\,cm in the transverse plane.
Particle identification~(PID) is performed using the specific ionization energy loss and time of flight information, and the resultant likelihood for protons is required to be greater than those for pions and kaons. Photon candidates are chosen from isolated clusters in the electromagnetic calorimeter~(EMC). Their energies are required to be greater than 25 MeV in the barrel ($\vert\!\cos\theta\vert<0.8$) region and 50 MeV in the end-cap ($0.86<\vert\!\cos\theta\vert<0.92$) region. Reconstructed clusters due to electronic noise or beam backgrounds are suppressed by requiring the EMC timing to be within [0, 700] ns after the event start time. To suppress background photons produced by hadronic interactions in the EMC, and secondary photons from bremsstrahlung radiation, clusters within cone angles of $20^\circ$ and $30^\circ$ around the extrapolated positions in the EMC of protons and anti-protons, respectively, are rejected~\cite{BESIII:2024gjj}. At least one photon candidate is required for further analysis. 

A four constraint (4C) kinematic fit, requiring energy and momentum conservation between the initial and final states, is imposed under the hypothesis of $e^+e^-\to J/\psi\to\gamma p\bar{p}$. If there is more than one combination due to multiple photon candidates, the combination with the minimum $\chi^2_{\rm 4C}$ is selected. The $\chi^2_{\rm 4C}$ is required to be less than 23 based on the optimization of the Figure-of-Merit defined as $\frac{S}{\sqrt{S+B}}$~\cite{Punzi:2003bu}, where $S$ in the numerator is the signal yield from MC simulation and $S+B$ in the denominator is the number of events from the data sample. Only candidates within the $\eta_c$ mass region are kept for the amplitude analysis.

After applying all selection criteria mentioned above, 479,652 candidate events for $J/\psi\to\gamma p\bar{p}$ survive in the data sample. The $J/\psi$ inclusive MC simulation contains two dominant background components: $J/\psi\to p\bar{p}\gamma^{\rm F}$ and $J/\psi\to p\bar{p}\pi^0$, where $\gamma^{\rm F}$ is a FSR photon. The process $J/\psi\to p\bar{p}\gamma^{\rm F}$ is well simulated in the inclusive MC sample with {\sc photos}~\cite{photos2}, which contributes to the background at a level of 22.8\% of the total events. The consistency between data and MC simulation is checked with the control sample $J/\psi\to p\bar{p}$, where the energy spectra of FSR photons show good agreement. Another exclusive MC sample is simulated for the $J/\psi\to p\bar{p}\pi^0$ decay based on a preliminary amplitude analysis result, and the corresponding background level is estimated to be 4.2\%. The contribution from other $J/\psi$ background processes is predicted to be less than 0.2\% and modeled with inclusive MC simulation in the amplitude fit. The non-$J/\psi$ background is studied with the data sample taken at $\sqrt{s}=3.080$ GeV, corresponding to an integrated luminosity of $(167.4\pm0.1)$~pb$^{-1}$~\cite{BESIII:2021cxx}. Its fraction, after taking into account the difference in luminosities~\cite{BESIII:2021cxx}, is estimated to be less than 0.7\% and ignored in further analysis.

\begin{figure*}[htbp]
	\centering
	\setlength{\abovecaptionskip}{0.cm}
	\includegraphics[width=18.5cm]{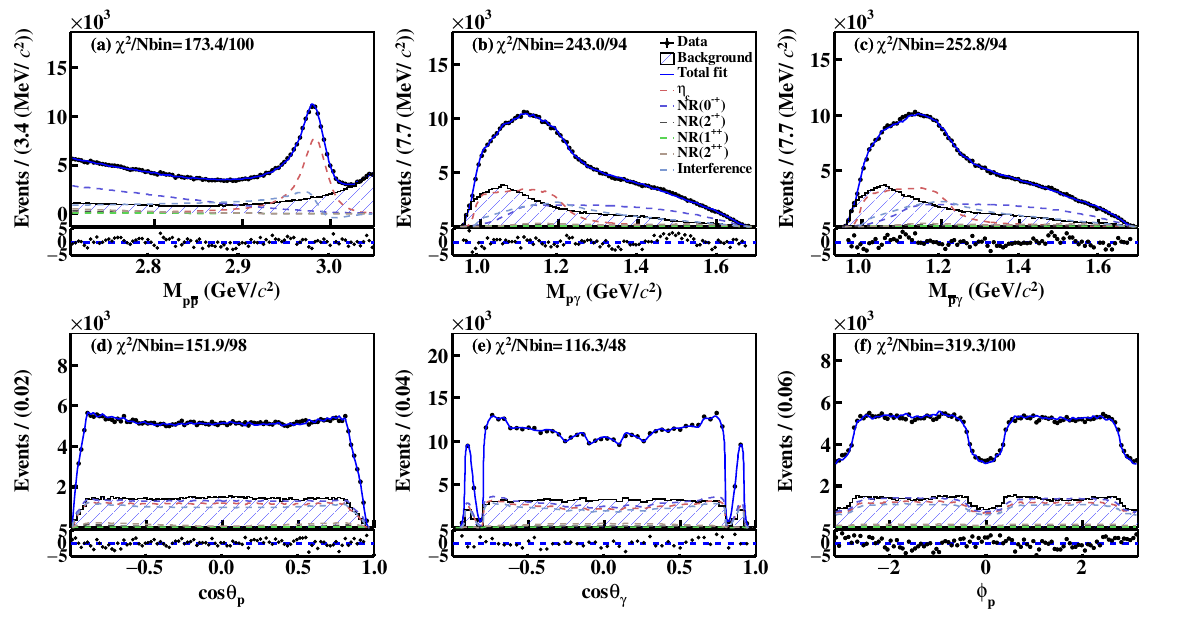}
	\caption{Projections of the amplitude analysis result on (a)~$M_{p\bar{p}}$, (b)~$M_{p\gamma}$, (c)~$M_{\bar{p}\gamma}$, and (d)~$\cos\theta_p$, (e)~$\cos\theta_{\gamma}$, and (f)~$\phi_p$. The dots with error bars are data, the blue solid curves are the total fit results, the dashed curves are various contributions, and the blue hatched histograms are the simulated backgrounds. The bottom sub-figures show the residuals ${(N_{\rm dt}-N_{\rm fit})}/{\sqrt{N_{\rm dt}}}$, where $N_{\rm dt}$ and $N_{\rm fit}$ are the number of events of data, and the fit results, respectively. Here, the dips in (f)~$\phi_p$ are caused by the efficiency loss due to the requirement on polar angle of charged tracks $|\cos\theta|<0.93$.}
	\label{fig:fit_projection}
\end{figure*}

Figure~\ref{fig:fit_projection} shows the distributions of the $p\bar{p}$, $p\gamma$, and $\bar{p}\gamma$ invariant masses, as well as those of $\cos(\theta_\gamma)$, $\cos(\theta_p)$ and $\phi_p$. Here, $\theta_{\gamma}$ is the polar angle of the photon in the $J/\psi$ rest frame with the $z$-axis defined as the direction of the $e^+$ beam and $(\theta_{p},\phi_{p})$ are the polar and azimuthal angles, respectively, in the $p\bar{p}$ helicity frame.
No significant structures in the $M_{p\bar{p}}$ or $M_{p\gamma}$ two-body invariant mass spectra are seen, other than the $\eta_c$ meson. As a check, potential contributions from $N^*(1440/1520/1535/1650)\to\gamma p$ are estimated with
\begin{equation}
    N_{J/\psi}\cdot\mathcal{B}[J/\psi\to N^*\bar{p},~N^*\to p\pi^0+c.c.]\cdot\frac{\mathcal{B}(N^*\to p\gamma)}{\mathcal{B}(N^*\to p\pi^0)}\cdot\varepsilon,
\end{equation}
where $\mathcal{B}[J/\psi\to N^*\bar{p},~N^*\to p\pi^0+c.c.]$ is taken from Ref.~\cite{BES:2009ufh}; $\mathcal{B}(N^*\to p\gamma/\pi^0)$ is from the PDG~\cite{ParticleDataGroup:2024cfk}; and $\varepsilon$ is the detection efficiency of $J/\psi\to N^*\bar{p}~(N^*\to p\gamma)+c.c.$ determined using MC simulation. The sum of background fractions of $N^*$ baryons is less than 1.4\%, and the $N^{*}$ background is ignored in the amplitude analysis.

The covariant tensor amplitude constructed in Ref.~\cite{Dulat:2005in} is applied in the amplitude analysis, which is expressed as
\begin{equation}
\begin{split}
&A^{(s)}=\psi_\mu(p,m_{J/\psi})e^*_{\nu}(q,m_{\gamma}) \\
&\times \psi_{\lambda_s}(p_p,S_p;p_{\bar{p}},S_{\bar{p}})\sum_{i}\Lambda_i U_i^{\mu\nu\lambda_s}.\\
\end{split}
\end{equation}
Here, $\psi_\mu(p,m_{J/\psi})$ is the polarization four-vector of the $J/\psi$ with a spin projection $m_{J/\psi}$ and four momentum $p$; $e_\nu(q,m_\gamma)$ is the polarization four-vector of the photon with spin projections $m_{\gamma}$ and four momentum $q$; $\psi_{\lambda_s}(p_p,S_p;p_{\bar{p}},S_{\bar{p}})$ is the spin wave function of the proton and anti-proton system with polarizations $S_{p,\bar{p}}$ and momenta $p_{p,\bar{p}}$, where the index $s$ is the total spin of the $p\bar{p}$ system; $U_i^{\mu\nu\lambda_s}$ is the $i$-th partial wave amplitude with a coupling strength determined by a complex parameter $\Lambda_i$. The form of the $\psi$ four-vector is detailed in Ref.~\cite{Dulat:2005in}.

Summing over the polarizations, the squared amplitude is given as
\begin{equation}
\begin{split}
&|\mathcal{M}|^2 = \frac{1}{2}\sum_{S_p,S_{\bar{p}}=\pm\frac{1}{2}}\sum_{m_J=\pm 1}\sum_{m_\gamma=\pm 1}|A^{(s)}|^2\\
&= -\frac{1}{2}\sum_{i,j}\Lambda_i\Lambda^*_{j}\sum^2_{\mu=1}U_{i}^{\mu\nu\lambda_s} g^{(\perp\perp)}_{\nu\nu^\prime}U_{j}^{*\mu\nu^\prime\lambda^\prime_s}\sum_{S_p,S_{\bar{p}}}\psi^*_{\lambda_s}\psi_{\lambda^\prime_s}, \\
\label{eq:cross_section}
\end{split}
\end{equation}
with $-g^{(\perp\perp)}_{\mu\nu}=\sum_{m_\gamma}{ e^*_{\mu}(q,m_{\gamma})e_\nu(q,m_{\gamma})}$~\cite{Dulat:2005in}. For $J/\psi\to\gamma$``$0^{-+}$"$\to\gamma p\bar{p}$, $U_{i}^{\mu\nu\lambda_s}=\epsilon^{\mu\nu\rho\sigma}p_{\mu}q_{\sigma}B_1(Q_b)R$, where $\epsilon^{\mu\nu\rho\sigma}$ is the Levi-Civita tensor, $B_1(Q_b)$ is the Blatt-Weisskopf barrier factor~\cite{VonHippel:1972fg} with angular momentum $L=1$ and $Q_{b}$ is the momentum of $X$ in $J/\psi\to\gamma X$ with $X=\eta_c$ or non-resonant, $R$ describes the line shape, and $i$ and $j$ are iterated over all possible processes. Explicit expressions of $U^{\mu\nu\lambda_s}$ for other spin-parity cases are available in Ref.~\cite{Dulat:2005in}.

The line shape $R$ of $\eta_c$ is described by a relativistic Breit-Wigner function $\frac{1}{M_{\eta_c}^2-M^2_{p\bar{p}}-i M_{\eta_c}\Gamma_{\eta_c}}$, where the mass $M_{\eta_c}$ and width $\Gamma_{\eta_c}$ of $\eta_c$ vary freely in the fit. Additionally, the Blatt-Weisskopf barrier factor in $J/\psi\to\gamma\eta_c$ is replaced by the square root of the damping factor $f_d$~\cite{CLEO:2008pln,Anashin:2010nr}, which is widely used to suppress the divergent long tail of $\eta_c$. Two well-known damping factors $e^{-E^2_\gamma/{8\beta^2}}$ of CLEO-c~\cite{CLEO:2008pln} with $\beta$ floating and $E^2_{\gamma 0}/{\left(E_{\gamma_0}E_\gamma+(E_\gamma-E_{\gamma_0})^2\right)}$ of KEDR~\cite{Anashin:2010nr}, are considered in this work. Here, $E_{\gamma}=\frac{M^2_{J/\psi}-M^2_{p\bar{p}}}{2 M_{J/\psi}}$ is the energy of the radiative photon and $E_{\gamma_0}$ is the photon energy under the
assumption $M_{p\bar{p}}=M_{\eta_c}$. To account for detector resolution effects, the product $R_{i}\times R^*_{j}$ in $|\mathcal{M}|^2$ is convolved with a Gaussian function $G(\delta_M,\sigma_M)$. The mass shift $\delta_M=(1.01\pm0.07)$ MeV/$c^2$ and resolution $\sigma_M=(3.93\pm0.05)$ MeV/$c^2$ are determined by studying the control sample $\psi(3686)\to\gamma\chi_{c1},\chi_{c1}\to p\bar{p}$. For the non-resonant contributions, $R$ is modeled by a constant.

The complex coupling constants $\Lambda_i$ and the resonance parameters of $\eta_c$ are determined with a maximum likelihood fit. The log-likelihood function is constructed as
\begin{equation}
\ln \mathcal{L} = \ln \mathcal{L}_{\rm dt} - \ln \mathcal{L}_{\rm bg},
\end{equation}
where $\ln \mathcal{L}_{\rm dt(bg)}$ sums over all the data or simulated background events $N_{\rm dt(bg)}$ and is defined as
\begin{equation}
\ln \mathcal{L}_{\rm dt(bg)} = \sum^{N_{\rm dt(bg)}}_{k=1}\ln\left[\frac{|\mathcal{M}(p^k)|^2}{\int\epsilon(p)|\mathcal{M}(p)|^2 \Phi_{3}(p)dp}\right],
\end{equation}
where $p$ is the momentum of the final state particles and $\Phi_{3}$ is the phase space factor. Integral of $\int\epsilon(p)|\mathcal{M}(p)|^2 \Phi_{3}(p)dp$ is calculated numerically using MC events as
\begin{equation}
\int\epsilon(p)|\mathcal{M}(p)|^2 \Phi_{3}(p)dp \propto \sum^{N_{\rm MC}}_{k_{\rm MC}=1}|\mathcal{M}(p^{k_{\rm MC}})|^2.
\end{equation}
Here, $N_{\rm MC}=2.3\times10^{6}$ is the number of phase space events that survive the data selection. The fit fractions with detection efficiency are estimated with
\begin{equation}
\small
f_{i} = \int{\epsilon(p)|\mathcal{M}(p)_{i}|^2 \Phi_{3}(p)dp} \bigg/ \int{\epsilon(p)|\mathcal{M}(p)|^2 \Phi_{3}(p)dp},
\end{equation}
where $\mathcal{M}_{i}$ is the amplitude of contribution $i$ alone.

First the CLEO-c and KEDR damping factors are tested with $\eta_c$ and all the potential NR contributions included. Since the KEDR damping factor results in a better log-likelihood value with a statistical significance $\sqrt{2\times\Delta\ln\mathcal{L}}=4.7\sigma$, as well as improved fit quality $(\Delta\chi^2/{\rm nbin}=10.4/100)$ in the $M_{p\bar{p}}$ projection compared to the CLEO-c form, the KEDR form is used in this analysis. An additional test is performed by excluding the damping factor, which yields a much worse fit quality with  $\Delta\chi^2/{\rm nbin}=49.1/100$ in the $M_{p\bar{p}}$ projection. All 13 potential NR contributions, each of which corresponds to different angular momentum and spin coupling in the transitions, are excluded from the solution one at a time, and the corresponding statistical significances are calculated based on the change of log-likelihood function $\Delta\ln\mathcal{L}$ and the number of free parameters $\Delta N_{\rm par}=2$. Five components with statistical significance greater than 3$\sigma$, including three waves with $J^{P}=$ $0^{-}$, $1^+$, and $2^+$, as well as two waves with $J^{P}=$ $2^-$, are kept in the final solution. By performing a scan on the phase angle $\phi$ between $\eta_c$ and $0^{-+}$ NR, two local minima are found. The best one exhibits a superior log-likelihood value with a statistical significance of $7.7\sigma$, hence we only consider it in the further analysis.

Figure~\ref{fig:fit_projection} shows the projections of the nominal amplitude analysis result. 
The mass and width of $\eta_c$ are determined to be $M_{\eta_c}=(2984.55\pm0.09_{\rm stat})$ MeV/$c^2$ and $\Gamma_{\eta_c}=(29.74\pm0.17_{\rm stat})$ MeV. The fit fraction of $\eta_c$ is determined to be $f_{\eta_c}=(30.94\pm0.24)\%$ and the product BF, $\mathcal{B}(J/\psi\to\gamma\eta_c)\times\mathcal{B}(\eta_c\to p\bar{p})=\frac{(N_{\rm dt}-N_{\rm bg})\cdot f_{\eta_c}}{\varepsilon_{\eta_c}\cdot N_{J/\psi}}$, is calculated to be $(2.11\pm0.02_{\rm stat})\times10^{-5}$. Here, $N_{\rm dt}-N_{\rm bg}$ is the net number of signal events, $N_{J/\psi}=(10.087\pm0.044)\times10^{9}$ is the number of $J/\psi$ events~\cite{BESIII:2021cxx}, and $\varepsilon_{\eta_c}=50.55\%$ is the signal efficiency determined with MC simulation based on the amplitude analysis result.

As a comparison, we have also tried to extract the $\eta_c$ signal yield via one-dimensional fit to the $M_{p\bar{p}}$ spectrum following previous publications. Without the interference between $\eta_c$ and NR contributions~\cite{CLEO:2008pln,Anashin:2010nr,Anashin:2014wva,Gaiser:1985ix}, the fit model can not provide an acceptable description of the data sample, and the fitted $\eta_c$ signal yield deviates from our nominal result by more than 30\%. With the interference considered following Ref.~\cite{BESIII:2024rdn}, two indistinguishable solutions with distinct interference patterns and $\eta_c$ signal yields are observed as expected. Additionally, the fraction of NR components other than $0^{-+}$ are found to be about 15\% in the amplitude analysis, which can not be determined in the one-dimensional fit and thereby causes large uncertainties~\cite{BESIII:2024rdn}. In contrast, the amplitude analysis successfully overcome these two issues.



\begin{table}[htbp]
	\centering
	\setlength\tabcolsep{5pt}
	\caption{Relative systematic uncertainties of the product BF $\mathcal{B}(J/\psi\to\gamma\eta_c)\times\mathcal{B}(\eta_c\to p\bar{p})$, and absolute systematic uncertainties of mass and width of $\eta_c$. The symbol ``$\star$" indicates negligible, and ``---" indicates not applicable.}
	\begin{tabular}{cccc}
		\hline\hline
		Source & $\mathcal{B}$~(\%) & $M_{\eta_c}$~(MeV/$c^2$) & $\Gamma_{\eta_c}$~(MeV) \\
		\hline
		$N_{J/\psi}$ & 0.5 & --- & --- \\
		Tracking & 0.2 & --- & --- \\
		PID & 0.3 & --- & --- \\
		Photon & 1.0 & --- & --- \\
		4C kinematic fit & 0.5 & --- & --- \\
		NR line shape & 0.4 & 0.38 & 0.14 \\
		Insig. NR waves & 1.8 & $\star$ & 0.13 \\
		Background & 0.8 & $\star$ & 0.06 \\
		Mass calibration & $\star$ & 0.37 & 0.24 \\
		Fit bias & 2.0 & 0.55 & 0.10 \\
		\hline
		Total & 3.1 & 0.77 & 0.33 \\
		\hline\hline
	\end{tabular}
	\label{tab:sys_sum}
\end{table}

The systematic uncertainties on the product $\mathcal{B}(J/\psi\to\gamma\eta_c)\times\mathcal{B}(\eta_c\to p\bar{p})$ and the resonance parameters of $\eta_c$ are summarized in Table~\ref{tab:sys_sum}. The uncertainty on the total number of $J/\psi$ events is 0.5\%~\cite{BESIII:2021cxx}. The systematic uncertainties due to the tracking and PID of $p(\bar{p})$ are studied using the control sample $J/\psi\to p\bar{p}\pi^+\pi^-$, and are determined to be 0.2\% and 0.3\%, respectively. The systematic uncertainty for photon reconstruction is assigned to be 1.0\%~\cite{BESIII:2011ysp}. The systematic uncertainty associated with the 4C kinematic fit is estimated by performing corrections on the charged track helix parameters in the MC simulation. The difference between the detection efficiencies obtained with and without the helix parameter correction~\cite{BESIII:2012mpj} is taken as the systematic uncertainty.

The systematic uncertainty due to the NR line shape is estimated by modeling the ${\rm NR}(0^{-+})$ component with an alternative line shape, the magnitude and phase of which are allowed to vary linearly as a function of $M_{p\bar{p}}$. The systematic uncertainty from NR waves considered insignificant is estimated by including all 13 potential waves. The systematic uncertainty in the background estimation is studied by varying the contribution from $J/\psi\to p\bar{p}\pi^0$ within the uncertainty of the quoted BF, and by including the simulated $N^*\to p\gamma$ contribution as background. The systematic uncertainty due to the mass calibration parameters, mass shift $\delta_M$ and resolution $\sigma_M$, is estimated by varying them within their statistical uncertainties and by using the control sample $\psi(3686)\to\gamma\chi_{c2},\chi_{c2}\to p\bar{p}$ as an alternative. For all four sources listed above in this paragraph, the amplitude fit is reperformed, and the largest difference for each source is assigned as its systematic uncertainty. The systematic uncertainty caused by a possible fit bias is studied by performing input and output checks with toy MC samples. The difference between the input and the averaged output values, predominantly attributed to detector effects and the statistical fluctuations of the phase space MC sample used in amplitude fit, is conservatively assigned as this uncertainty.


In summary, the first amplitude analysis of $J/\psi\to\gamma p\bar{p}$ with $M_{p\bar{p}}$ in the $\eta_c$ mass region $[2.70,3.05]$~GeV/$c^2$ is performed. Compared to the one-dimensional fits to the $\eta_c$ spectrum~\cite{CLEO:2008pln,Anashin:2010nr,Anashin:2014wva,Gaiser:1985ix,BESIII:2024rdn}, our amplitude analysis approach provides more reliable estimations of the different non-resonant components, and their interference with $\eta_c$, hence, avoids several 10\%–level biases and uncertainties. The mass and width of $\eta_c$ are measured to be $(2984.55\pm0.09_{\rm stat}\pm0.77_{\rm syst})$ MeV/$c^2$ and $(29.74\pm0.17_{\rm stat}\pm0.33_{\rm syst})$ MeV, respectively, which are in good agreement with the PDG world averages~\cite{ParticleDataGroup:2024cfk}. The product BF $\mathcal{B}(J/\psi\to\gamma\eta_c)\times\mathcal{B}(\eta_c\to p\bar{p})$ is determined to be $(2.11\pm0.02_{\rm stat}\pm0.07_{\rm syst})\times10^{-5}$, whose precision is improved by one order of magnitude compared to the previous measurements~\cite{ParticleDataGroup:2024cfk}. 

\begin{figure}[htbp]
	\centering
	\setlength{\abovecaptionskip}{0.0cm}
	\includegraphics[width=9cm]{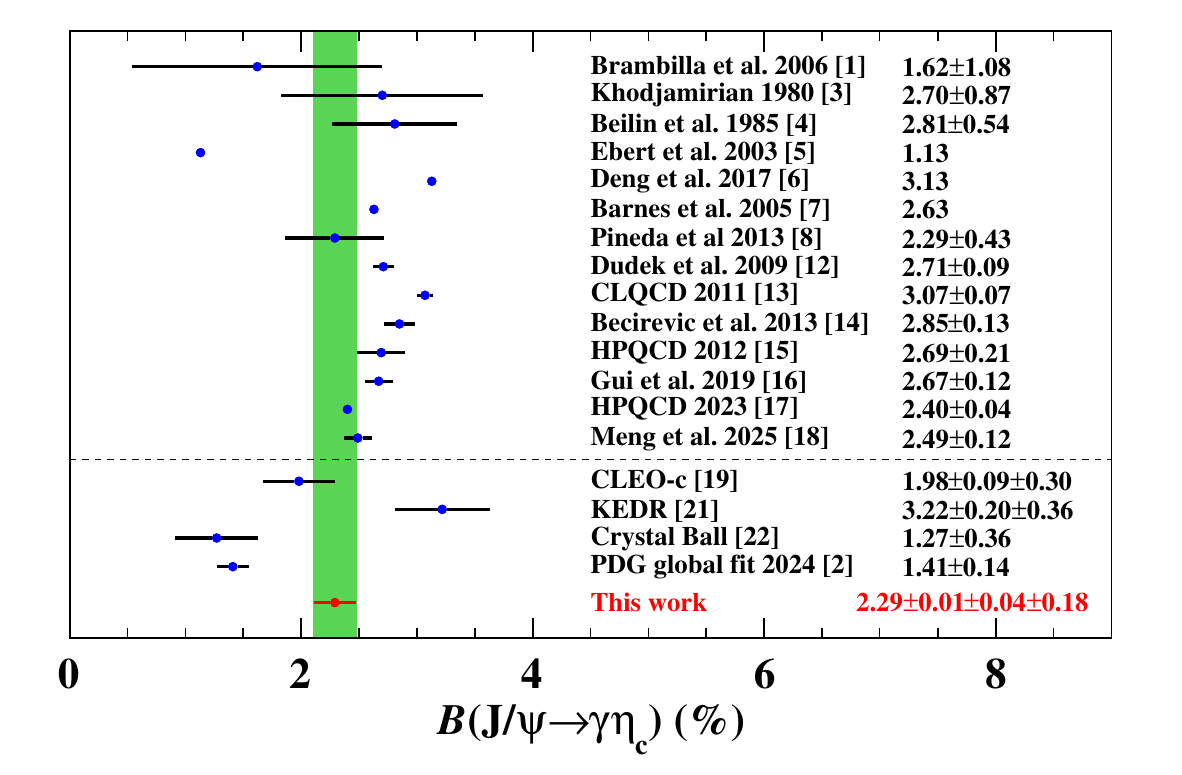}
	\caption{Comparison of $\mathcal{B}(J/\psi\to\gamma\eta_c)$ derived in this work with theoretical calculations~\cite{Khodjamirian:1979fa,Beilin:1984pf,Ebert:2002pp,Deng:2016stx,Barnes:2005pb,Brambilla:2005zw,Pineda:2013lta,Segovia:2021bjb,Guo:2019xqa,Dudek:2006ej,Dudek:2009kk,Chen:2011kpa,Becirevic:2012dc,Donald:2012ga,Gui:2019dtm,Colquhoun:2023zbc,Meng:2024axn}, other experimental measurements~\cite{CLEO:2008pln,Anashin:2010nr,Anashin:2014wva,Gaiser:1985ix}, and the PDG global fit~\cite{ParticleDataGroup:2024cfk}. The green band corresponds to the $\pm1\sigma$ region of $\mathcal{B}(J/\psi\to\gamma\eta_c)$ in this work, and the uncertainties are statistical, systemtic, and those from the other product BFs used in the calculation.}
	\label{fig:BFcompare}
\end{figure}

\begin{figure}[htbp]
	\centering
	\setlength{\abovecaptionskip}{0.0cm}
	\includegraphics[width=9cm]{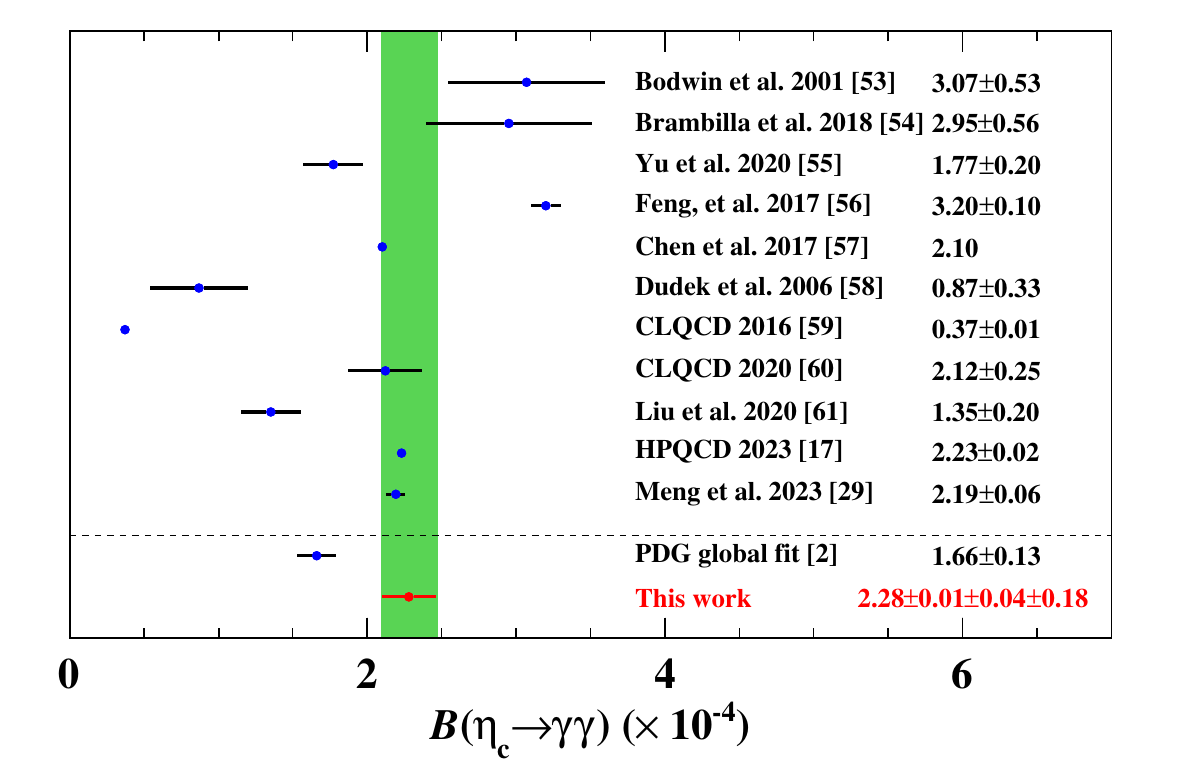}
	\caption{Comparison of $\mathcal{B}(\eta_c\to\gamma\gamma)$ derived in this work with theoretical calculations~\cite{CLQCD:2020njc,CLQCD:2016ugl,Liu:2020qfz,Bodwin:2001pt,Brambilla:2018tyu,Yu:2019mce,Dudek:2006ut,Chen:2016bpj,Feng:2017hlu,Colquhoun:2023zbc,Meng:2021ecs} and the PDG global fit~\cite{ParticleDataGroup:2024cfk}. The green band corresponds to the $\pm1\sigma$ region of $\mathcal{B}(\eta_c\to\gamma\gamma)$ in this work, and the uncertainties are statistical, systemtic, and those from the other product BFs used in the calculation.}
	\label{fig:BFcompare_gamgam}
\end{figure}

Given most measurements used in the PDG global fits ignored interference effects~\cite{ParticleDataGroup:2024cfk}, the PDG fitted $\mathcal{B}(\eta_c\to p\bar{p})$ is not quoted for the extraction of $\mathcal{B}(J/\psi\to\gamma\eta_c)$. Instead, two unique processes $J/\psi\to\gamma\eta_c,\eta_c\to\gamma\gamma$ and $\gamma\gamma\leftrightarrow p\bar{p}$ around $\eta_c$ peak, which benefit from limited interference effects~\cite{BESIII:2024rex,E760:1995rep}, are quoted. Combining this result with the products $\mathcal{B}(\eta_c\to p\bar{p})\times\mathcal{B}(\eta_c\to \gamma\gamma)=(2.1\pm0.3)\times10^{-7}$ averaged based on Refs.~\cite{Belle:2005fji,FermilabE835:2003ula,E760:1995rep} and the recently reported $\mathcal{B}(J/\psi\to\gamma\eta_c)\times\mathcal{B}(\eta_c\to \gamma\gamma)=(5.23\pm0.40)\times10^{-6}$~\cite{BESIII:2024rex}, we obtain
\begin{widetext}
\begin{equation}
\small
\begin{split}
\mathcal{B}(J/\psi\to\gamma\eta_{c})&=\sqrt{\frac{\left[\mathcal{B}(J/\psi\to\gamma\eta_c)\times\mathcal{B}(\eta_c\to p\bar{p})\right]\times \left[\mathcal{B}(J/\psi\to\gamma\eta_c)\times\mathcal{B}(\eta_c\to \gamma\gamma)\right]}{\left[\mathcal{B}(\eta_c\to p\bar{p})\times\mathcal{B}(\eta_c\to \gamma\gamma)\right]}}=(2.29\pm0.01\pm0.04\pm0.18)\%,\\
\mathcal{B}(\eta_{c}\to\gamma\gamma)&=\sqrt{\frac{\left[\mathcal{B}(\eta_c\to p\bar{p})\times\mathcal{B}(\eta_c\to \gamma\gamma)\right]\times\left[\mathcal{B}(J/\psi\to\gamma\eta_c)\times\mathcal{B}(\eta_c\to \gamma\gamma)\right]}{\left[\mathcal{B}(J/\psi\to\gamma\eta_c)\times\mathcal{B}(\eta_c\to p\bar{p})\right]}}=(2.28\pm0.01\pm0.04\pm0.18)\times10^{-4},\\
\mathcal{B}(\eta_{c}\to p\bar{p})&=\sqrt{\frac{\left[\mathcal{B}(\eta_c\to p\bar{p})\times\mathcal{B}(\eta_c\to \gamma\gamma)\right]\times\left[\mathcal{B}(J/\psi\to\gamma\eta_c)\times\mathcal{B}(\eta_c\to p\bar{p})\right]}{\left[\mathcal{B}(J/\psi\to\gamma\eta_c)\times\mathcal{B}(\eta_c\to \gamma\gamma)\right]}}=(0.92\pm0.01\pm0.02\pm0.07)\times10^{-3},\\
\end{split}
\end{equation}
\end{widetext}
where the uncertainties are statistical, systematic, and those from the other product BFs used in the calculation. Some systematic uncertainties are correlated between our work and Ref.~\cite{BESIII:2024rex}, which are dominated by the photon detection and the form of the damping factor. Figures~\ref{fig:BFcompare} and~\ref{fig:BFcompare_gamgam} show comparisons of $\mathcal{B}(J/\psi\to\gamma\eta_c)$ and $\mathcal{B}(\eta_{c}\to\gamma\gamma)$ determined in this study, with various theoretical calculations and other measurements. Our results deviate from the PDG global fit values~\cite{ParticleDataGroup:2024cfk} by $3\sigma$, but are in good agreement with the latest LQCD calculations~\cite{Colquhoun:2023zbc,Meng:2021ecs,Meng:2024axn}. This helps resolve a long-standing puzzle. Furthermore, the obtained BF $\mathcal{B}(\eta_c\to p\bar{p})$ deviates from the PDG global fit values by 3$\sigma$. Additionally, using $M_{J/\psi}=3096.9$ MeV/$c^2$~\cite{ParticleDataGroup:2024cfk}, the mass splitting between the $J/\psi$ and $\eta_c$ $c\bar{c}$ systems is determined to be $(112.35\pm0.77)$ MeV/$c^2$, which is consistent with the results calculated in Refs.~\cite{Follana:2006rc,Burch:2009az,Kawanai:2011jt,Hatton:2020qhk}.

The BESIII Collaboration thanks the staff of BEPCII (https://cstr.cn/31109.02.BEPC) and the IHEP computing center for their strong support. This work is supported in part by National Key R\&D Program of China under Contracts Nos. 2025YFA1613900, 2023YFA1606000, 2023YFA1606704; National Natural Science Foundation of China (NSFC) under Contracts Nos. 11635010, 11935015, 11935016, 11935018, 12025502, 12035009, 12035013, 12061131003, 12192260, 12192261, 12192262, 12192263, 12192264, 12192265, 12221005, 12225509, 12235017, 12361141819; the Chinese Academy of Sciences (CAS) Large-Scale Scientific Facility Program; CAS under Contract No. YSBR-101; 100 Talents Program of CAS; The Institute of Nuclear and Particle Physics (INPAC) and Shanghai Key Laboratory for Particle Physics and Cosmology; German Research Foundation DFG under Contract No. FOR5327; Istituto Nazionale di Fisica Nucleare, Italy; Knut and Alice Wallenberg Foundation under Contracts Nos. 2021.0174, 2021.0299; Ministry of Development of Turkey under Contract No. DPT2006K-120470; National Research Foundation of Korea under Contract No. NRF-2022R1A2C1092335; National Science and Technology fund of Mongolia; Polish National Science Centre under Contract No. 2024/53/B/ST2/00975; Swedish Research Council under Contract No. 2019.04595; U. S. Department of Energy under Contract No. DE-FG02-05ER41374

\end{document}